\documentclass[]{raa}
\usepackage{graphicx,times}
\usepackage{natbib}
\providecommand{\bm}[1]{\mbox{\boldmath$#1$\unboldmath}}

\begin{document}

\title{Star Formation Properties in Barred Galaxies (SFB). II. NGC 2903 and NGC 7080}

 \volnopage{ {\bf 0000} Vol.\ {\bf 0} No. {\bf XX}, 000--000}
   \setcounter{page}{1}

\author{Zhi-Min Zhou \inst{1,2,3}
 \and Chen Cao \inst{4,5}
 \and Hong Wu \inst{1,3}}

   \institute{National Astronomical Observatories, Chinese Academy of Sciences, Beijing 100012, China;
              {\it zmzhou@bao.ac.cn}\\
        \and Graduate School, Chinese Academy of Sciences, Beijing 100039, China\\
        \and Key Laboratory of Optical Astronomy, National Astronomical Observatories, Chinese Academy of Sciences, Beijing 100012, China\\
        \and School of Space Science and Physics, Shandong University at Weihai, Weihai, Shandong 264209, China\\
        \and Shandong Provincial Key Laboratory of Optical Astronomy \& Solar-Terrestrial Environment, Weihai, Shandong, 264209, China\\
\vs \no
   {\small Received [year] [month] [day]; accepted [year] [month] [day] }
}

%% Abstract
\abstract{Stellar bars are important for the secular evolution of disk galaxies because they can drive gas into the galactic central regions. To investigate the star formation properties in barred galaxies, we presented a multi-wavelength study of two barred galaxies NGC 2903 and NGC 7080. We performed the three-component bulge-disk-bar decomposition using the 3.6 $\mu$m images, and identified the bulges in the two galaxies as pseudobulges. Based on the narrowband H$\alpha$ images, the star formation clumps were identified and analyzed. the clumps in the bulge regions have the highest star formation rate surface densities in both galaxies, while the star formation activities in the bar of NGC 2903 are more intense than those in the bar of NGC 7080. Finally, we compared with the scenario of bar-driven secular evolution in previous studies, and discussed the possible evolutionary stages of the two galaxies.
}

\authorrunning{Zhou et al.}
\titlerunning{Multi-wavelength Study of NGC~2903 and NGC 7080}
\maketitle

%%Sect. 1 Introduction
\section{Introduction}

The Universe is in transition, in which galactic evolution is dominated from an early universe by hierarchical clustering to a future by secular processes. Galaxy secular evolution is the slow rearrangement of energy and mass driven by collective phenomena such as bars, oval disks, spiral structures, and triaxial dark halos in galaxy disks \citep{KK04}. One important consequence of secular processes is the buildup of dense central components in disk galaxies, i.e., pseudobulges. Different from classical, merger-built bulges, pseudobulges are made slowly out of disk gas, and are similar in many ways to disk galaxies \citep{Kormendy05, Fisher06, Gadotti09}.

Galactic large-scale bars are the important internal drivers for secular evolution. It is well known that more than 70\% of normal bright galaxies are barred in the local universe \citep[e.g.,][]{Knapen99, Eskridge00}. The fraction of strong bars may remain fairly constant ($\sim$ 30\%) out to z $\sim$ 1 \citep{Jogee04}, and it is still about 10\% if it really declines with redshift \citep{Sheth08}. Bars have important effect on the redistribution of material in disks. A connection between bar dynamics and star formation has been found \citep{Martinez11}. The nonaxisymmetry in the gravitational potential of bars can rearrange disk gas, and lead to the inflow of gas from the outer disk to the central regions, which can trigger starbursts \citep{Sellwood93, Athanassoula03, Sheth05}. Observations have indicated that more molecular gas is concentrated in the central several kiloparsecs (kpc) of barred spirals than unbarred disks \citep{Sakamoto99, Helfer03, Sheth05, Kormendy05}. The enhanced star formation activities are also found in the central regions of the barred disks \citep{Ho97, Fisher06, Regan06}.

In a previous work of this series \citep[][hereafter Paper I]{Zhou11}, we have presented a study of star formation properties in the barred galaxy NGC 7479, which is is possibly in an early stage of the secular evolution triggered by the stellar bar at present. With the aim of better understanding the observational properties of bar-driven secular evolution, in this paper, we describe a multi-wavelength study of two isolated barred galaxies NGC 2903 and NGC 7080 (Messier 66). NGC 2903 is a SBd galaxy showing a symmetric strong bar, and is isolated from large companions, preventing major merger effects in the results. It located at a distance of $\sim$ 8.9 Mpc, corresponding to a scale at the galaxy of 56 pc arcsec$^{-1}$ \citep{Drozdovsky00}. NGC 7080 is a SBb galaxy \citep{de91}, located at a distance of 62.7 Mpc, corresponding to a scale of 295 pc arcsec$^{-1}$. The significant property of this galaxy is the intense star formation activity in its bulge \citep{Filippenko85}. The basic parameters of the NGC 2903 and NGC 7080 are given in Table~\ref{Table1}.

This paper is outlined as follows: In Section 2 we describe the imaging observations and the archival data. We present the methodology to analyze the structure properties and list the results in Section 3. The following section contains the main results regarding the properties of star formation activities in the two galaxies. Section 5 discusses the effect of bars to the secular evolution of galaxies. A summary is given in Section 6.

%%Sect.2 DATA%%
\section{OBSERVATIONS AND DATA REDUCTION}

\subsection{Optical Imaging}
Optical images of NGC 2903 and 7080 were obtained with the 2.16m telescope at Xinglong Observatory of the National Astronomical Observatories of the Chinese Academy of Sciences\footnote{http://www.xinglong-naoc.org/English/216.html} on 2009 December 22 and 2006 September 24, respectively. The observations were carried out by the BAO Faint Object Spectrograph and Camera with a 2048 $\times$ 2048 Loral Lick CCD with a pixel size of 0.305$''$, giving a field of view of about 10$'$ $\times$ 10$'$. Both galaxies were imaged with the broadband R filter and the red-shifted H$\alpha$ filters. For NGC 2903, an exposure of 600 s was made in the broadband R filter, along with an 3000 s image in the narrowband H$\alpha$ filter centered at 6562 \AA~with a FWHM of 70 \AA. For NGC 7080, the exposure time was 600 s for the broadband R filter, and was 2400 s for the narrowband H$\alpha$ filter, which was centered at 6660 \AA~with a FWHM of 70 \AA. The two narrowband filters contain both H$\alpha$ and the [N {\sc ii}] $\lambda\lambda$6563, 6583 emission lines.

The observed data were reduced in the standard way using the Image Reduction and Analysis Facility (IRAF\footnote{IRAF is the Image Reduction and Analysis Facility written and supported by the IRAF programming group at the National Optical Astronomy Observatories (NOAO) in Tucson, Arizona which is operated by AURA, Inc. under cooperative agreement with the National Science Foundation.}) software. First, all images were processed by overscan and bias subtractions, flat-field correction, cosmic-ray removal, astrometry, and flux calibration as in Paper I. Second, continuum subtraction from the H$\alpha$ filters was accomplished using the scaled R-band image. We refer the reader to Paper I for further details concerning the data and photometric reductions.

\subsection{IR Imaging}
Broadband mid-infrared images of NGC 2903 and NGC 7080 were acquired with the Infrared Array Camera \citep[IRAC;][]{Fazio04a} and Multiband Imaging Photometer for {\it Spitzer} \citep[MIPS;][]{Rieke04} on board {\it Spitzer Space Telescope} \citep{Werner04}. For NGC 2903, which was observed by {\it Spitzer} as part of the Local Volume Legacy Survey \citep[LVL;][]{Dale09}, we archived its images of the four IRAC (3.6, 4.5, 5.8, and 8 $\mu$m) and MIPS 24 $\mu$m bands from LVL database. For NGC 7080, which was observed by {\it Spitzer} as part of the Mid-IR Hubble Atlas of Galaxies \citep{Fazio04b}, we mosaicked its Basic Calibrated Data (BCD) images using MOPEX \citep[MOsaicker and Point source Extractor;][]{Makovoz05} version 18.1.5, and obtained the images with the same IR bands and pixel scales of NGC 2903. For both galaxies the FWHMs of the point-spread functions (PSF) at the four IRAC bands are $\sim$ 2$''$, along with the plate scales of 0.75$''$ pixel$^{-1}$, and the FWHMs of PSF at MIPS 24 $\mu$m band are $\sim$ 6$''$, along with the plate scales of 1.5$''$ pixel$^{-1}$.
For IRAC 8.0 $\mu$m band images, the corresponding scaled IRAC 3.6 $\mu$m band images were used to subtract the contribution from stellar continuum, and the scale factor of 0.232 from \citet{Helou04} was adopted. Then, the continuum-subtracted 8.0 $\mu$m emission was denoted by 8 $\mu$m(dust).

%%Sect.3 Global and Structural Properties %%
\section{Global and Structural Properties}
\subsection{Multi-wavelength Images}
\begin{figure}
\center
\includegraphics[angle=0,scale=.8]{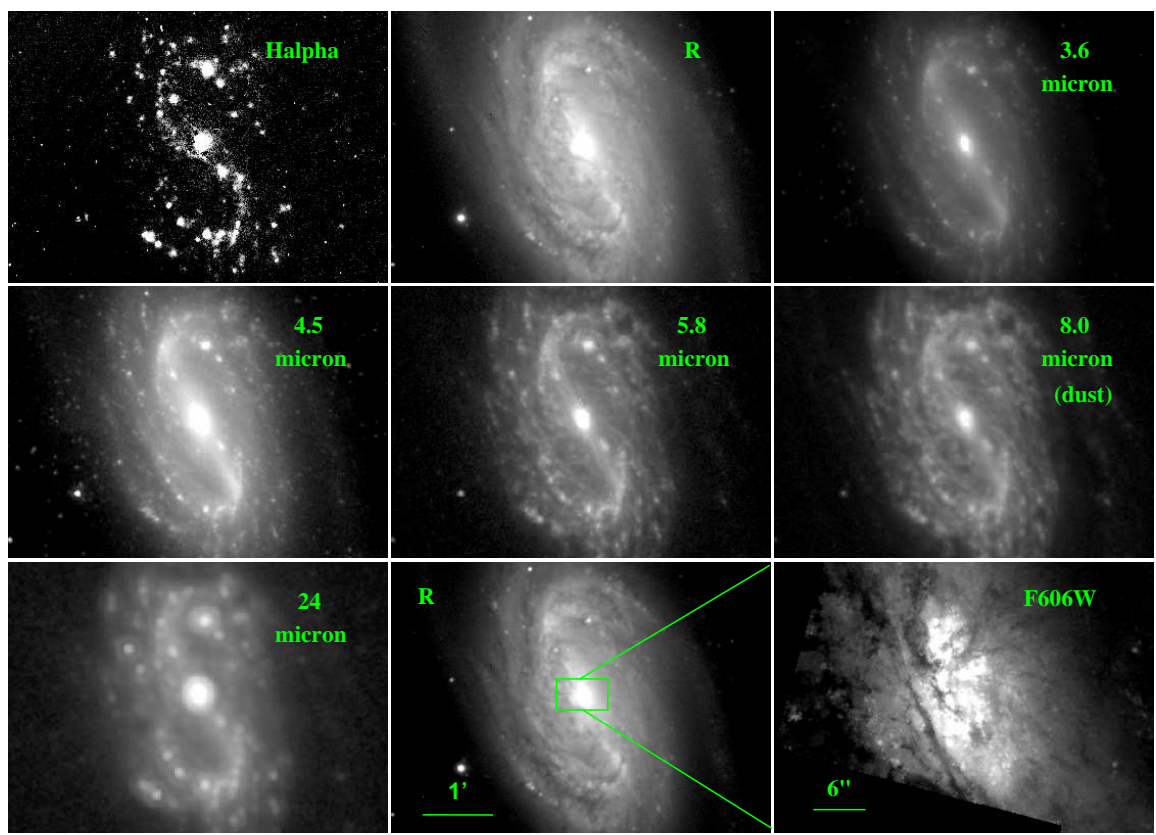}
\caption{A montage of optical and infrared images of NGC 2903. North is up and east is to the left. The H$\alpha$ and 8 $\mu$m (dust) images are continuum-subtracted using the scaled R-band image and scaled 3.6 $\mu$m image, respectively.  The HST WFPC2 F606W image in the last panel shows the central 20-arcsec region of NGC 2903.
\label{fig1}}
\end{figure}

Figure~\ref{fig1} shows the images of NGC 2903 in different spectral bands. In the optical R and IR 3.6 $\mu$m images, a stellar bar and a bright bulge-like feature are seen in the disk. A series of dusty spiral patterns are visible in the R image. These patterns are located around the bulge, and are more intense in the galactic western region. Several bright star-forming regions are clearly seen in the H$\alpha$ image. The brightest clump is located in the bulge region, indicating the intense star formation activity there. Other clumps are mainly located in two spiral arms, and also several clumps can be found in the bar region. In the images, H$\alpha$, 8 $\mu$m(dust), and 24 $\mu$m bands all can trace the star formation activities. The star-forming clumps in 8 $\mu$m(dust) and 24 $\mu$m images have similar distribution with those in H$\alpha$ image. Especially in the 24 $\mu$m image, the clumps are more pronounced. There is a position angle misalignment of $\sim$ 20\degr~ between the large-scale bar in H$\alpha$ image and in IR images, indicating the offset between the gaseous pattern and star formation in the bar. This offset was caused by the gas flow dependent star formation \citep{Sheth02}, and was also found in many other samples, e.g., M 101 \citep{Kenney91}, IC 342 \citep{Crosthwaite00}, NGC 3488 \citep{Cao08}.

\begin{figure}
\center
\includegraphics[angle=0,scale=.8]{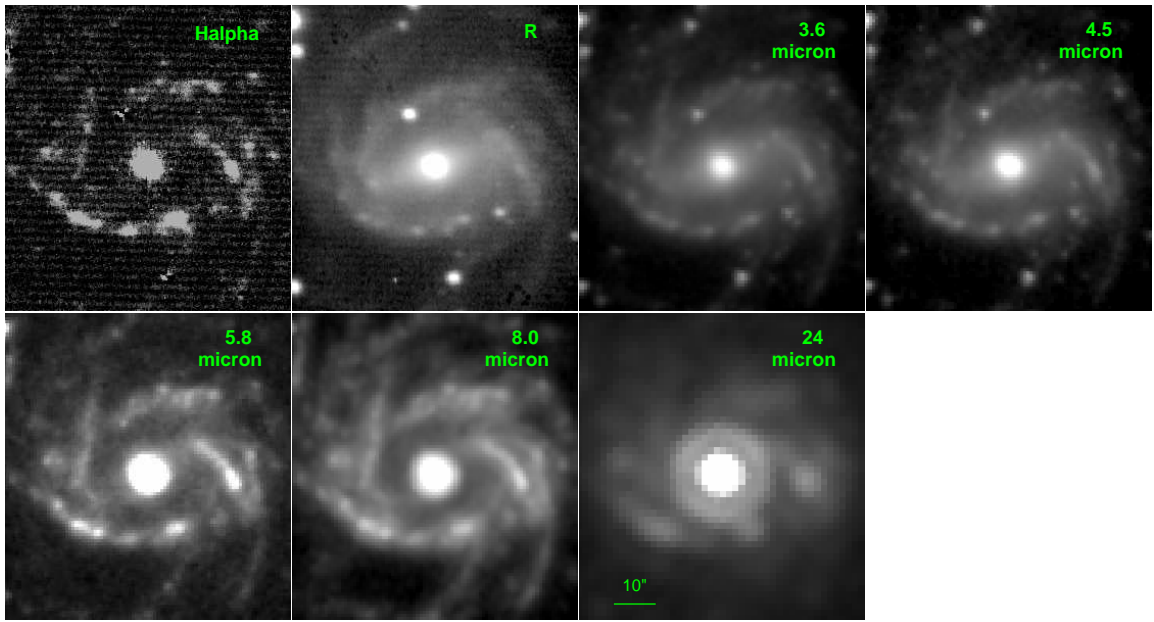}
\caption{Same as Figure~\ref{fig1}, but showing the multi-wavelength images of NGC 7080, and no images of the detailed inner region in this galaxy is shown.
\label{fig2}}
\end{figure}

A montage of the optical and infrared images of NGC 7080 is displayed in Figure~\ref{fig2}. NGC 7080 is a nearly face-on galaxy. Its galactic diameter is $\sim$ 32 kpc, similar with the $\sim$ 42 kpc of NGC2903. A bright bulge is clearly visible in each band image, especially in 5.8 $\mu$m, 8 $\mu$m(dust), and 24 $\mu$m images. A galactic bar is only present in R and near-IR images, and is less pronounced than that in NGC 2903. Comparing with NGC 2903, one important difference in NGC 7080 is that its on-going star formation is only detected at the galactic central region and along the spiral arms, but not on the galactic bar. In addition, the offset of the large-scale bar in different images is not clearly as NGC 2903.

\subsection{Image Decomposition and Structure Characterization}
In order to characterize galaxy structures, one useful way is to model the light distribution using analytic functions, because the structural properties of galaxy components, such as bulges, disks, and bars can be derived through the decomposition of the one-dimensional (1D) or two-dimensional (2D) light distribution. There are a series of algorithms developed for fitting these galactic structures. 1D decomposition was performed though fitting radial surface brightness profiles \citep[e.g.,][]{Kormendy77, Kent85}, and 2D decomposition was performed directly based on images using techniques such as GIM2D \citep{Simard02} and GALFIT \citep{Peng02}. In general, 2D image fitting is more advantageous than 1D profile fitting. In 2D methods, GIM2D is automated on bulge-disk decomposition in galaxy surveys, while GALFIT handles more compositions of structures using various brightness profiles.

The IRAC 3.6 $\mu$m band is an ideal tracer of the stellar compositions in nearby galaxies, although the polycyclic aromatic hydrocarbon (PAH) emission at 3.3 $\mu$m and young stars have some contributions \citep{Kendall11}. In the present work, we used the IRAC 3.6 $\mu$m images of our galaxies to perform 2D three-component bulge-disk-bar decomposition with GALFIT. GALFIT is a data analysis algorithm that adjust analytic functions to try and match the shape and profile of galaxies in digital images \citep{Peng02, Peng10}, and has been used in many previous works \citep[e.g.,][]{Weinzirl09, Orban11}. It is well known that the surface brightness profiles of disk and bulge can be described by an exponential function and a S\'ersic function, respectively. In addition, the bar can be well described by an elongated, low-index S\'ersic (n $<$ 1) profile \citep{Peng02}. Therefore, for each galaxy we iteratively fit three components: a S\'ersic bulge, an exponential disk and a S\'ersic bar.

\begin{figure}
\center
\includegraphics[angle=0,scale=.8]{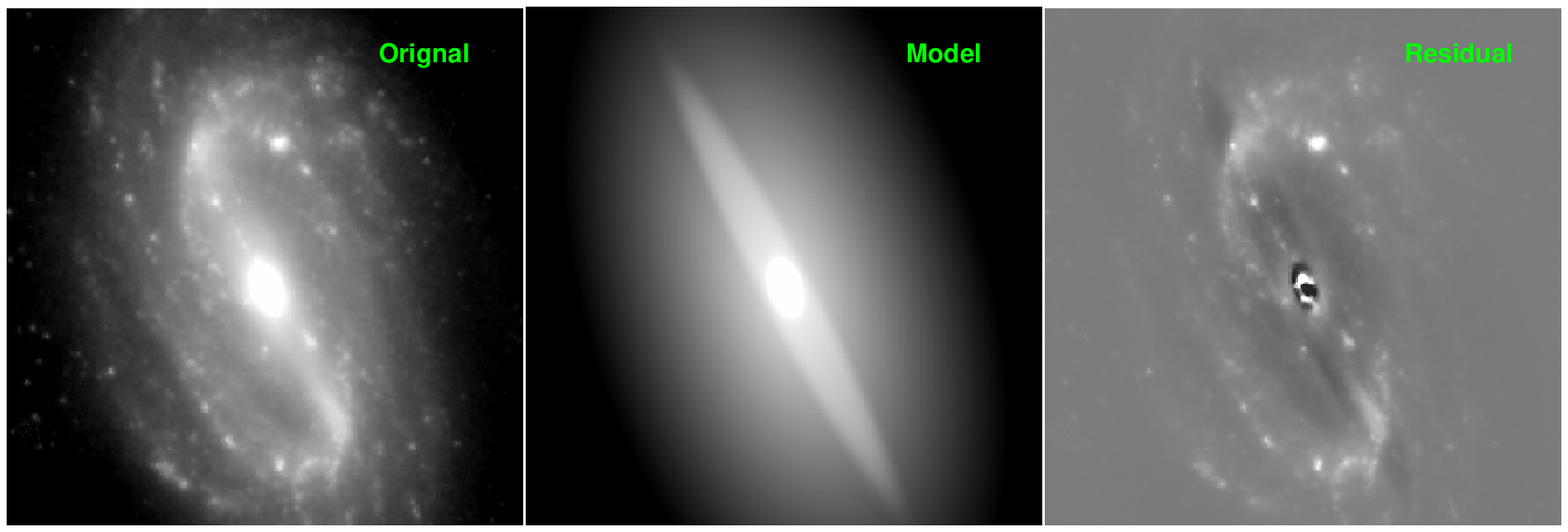}
\includegraphics[angle=0,scale=.8]{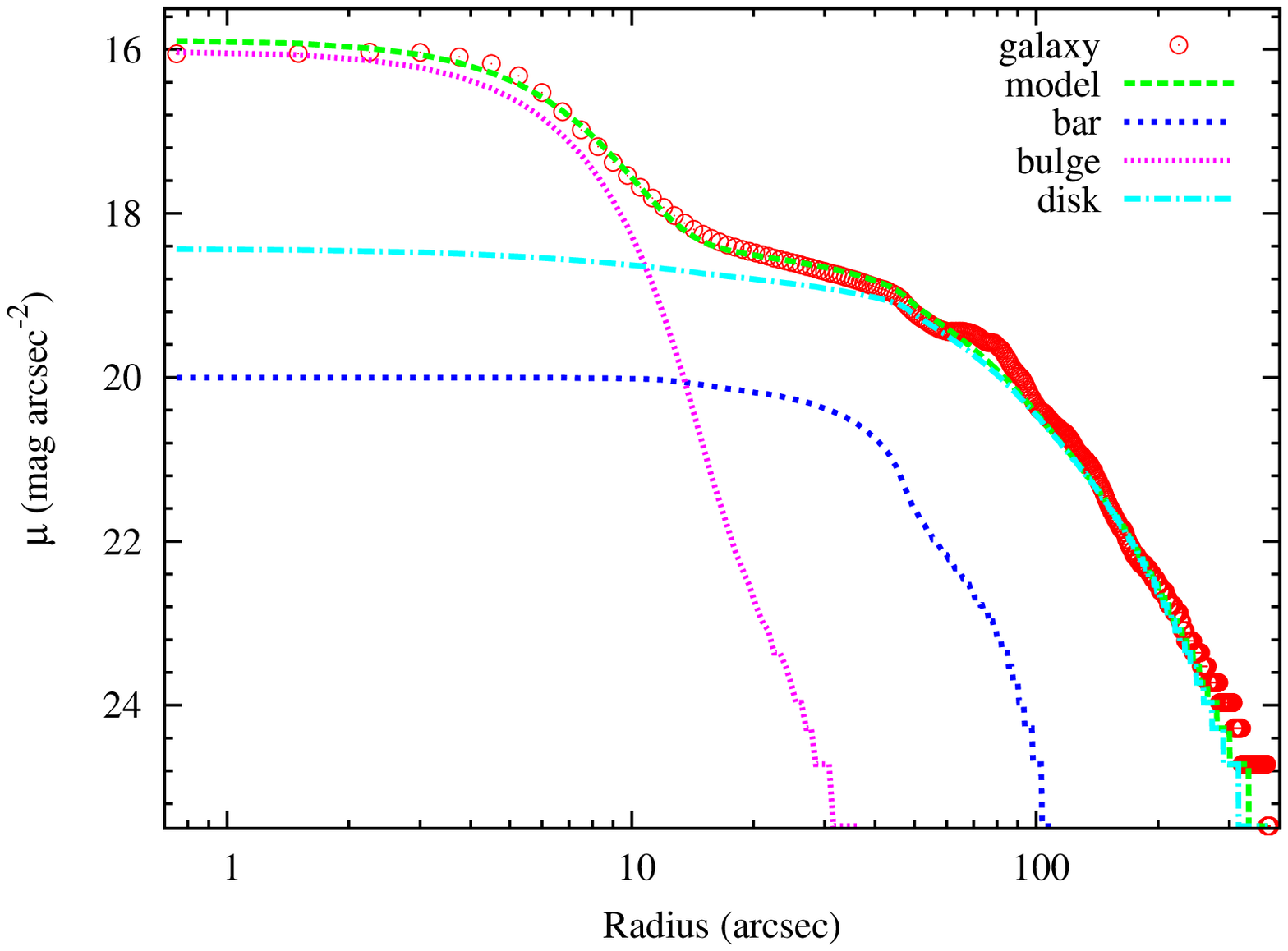}
\caption{Results of Bulge-Bar-Disk decomposition in the IRAC 3.6 $\mu$m band for NGC 2903. The images on the top show the galaxy of the original image, as well as the model and residual images. In the residual image, brighter shades indicate regions where the model is more luminous than the galaxy, whereas darker shades indicate regions where the model is fainter than the galaxy. The plots on the bottom show the surface brightness profiles of the galaxy and its model images, as well as individual components from the fitting results.
\label{fig3}}
\end{figure}

\begin{figure}
\center
\includegraphics[angle=0,scale=.8]{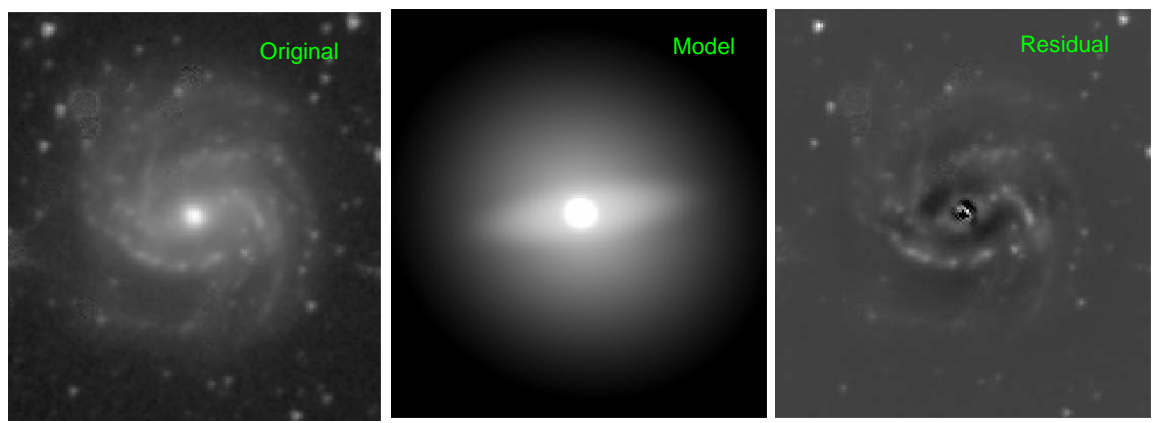}
\includegraphics[angle=0,scale=.8]{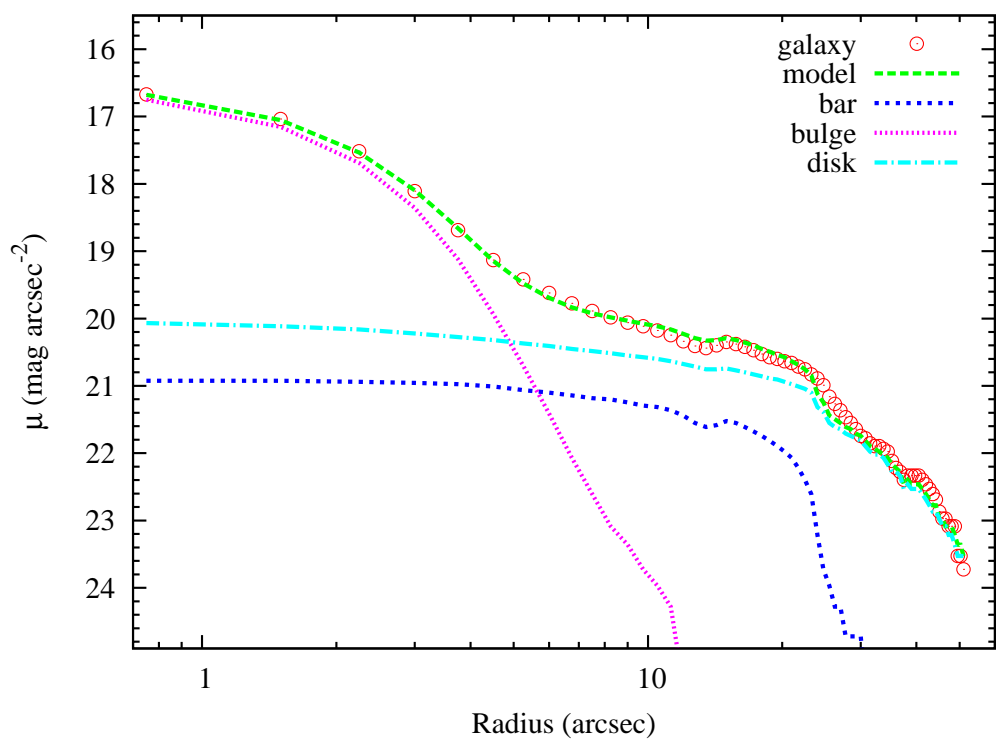}
\caption{Same as Figure~\ref{fig3}, but showing the Bulge-Bar-Disk decomposition of NGC 7080.
\label{fig4}}
\end{figure}

The relevant results of our decompositions are demonstrated in Figure~\ref{fig3} and Figure~\ref{fig4}, including the original image, an image of the model obtained with GALFIT, and a residual image of each galaxy. In the two figures, we also displayed surface brightness radial profiles of the galaxy, of each component in the model separately, and of the total model, for comparison. In the models generated with GALFIT, the S\'ersic indices of the bulges in the two galaxies are 0.50 and 0.68, respectively. The bulge-to-total luminosity ratios are both much smaller than 1/3. Because the candidate pseudobulge has some disk-like characteristics and has a S\'ersic index n $<$ 2, comparing with the classical bulge peak at n = 4 \citep{KK04, Fisher08, Fisher10}, the bulges in our two galaxies can be identified as the pseudobulges. The effective radius\footnote{The effective radius is the radius within which half of the total flux is contained.} of the bar in NGC 7080 is 5.21 kpc, longer than the 3.51 kpc of the bar in NGC 2903, and its relative size is twice of the bar in NGC 2903. While the ellipticity is 0.70 for the bar in NGC 7080, smaller than 0.88 of the bar in NGC 2903. In Table~\ref{Table2}, we summarized the model parameters from the fitting. 

Considering the the contaminant emission from hot dust and PAH in the integrated light at 3.6 $\mu$m images \citep{Meidt11}, we reduced the contamination in the 3.6 $\mu$m data following the method in \citet{Kendall11}, re-performed the decomposition using the corrected images. The results were also listed in Table~\ref{Table2}. Compared them with the original results, we found that the effect of PAHs in IRAC 3.6 $\mu$m could be negligible, especially for the parameters of the structures. Therefore, we still used the parameters firstly derived in the following text.

%%Sect.4 the Star formation properties %%
\section{Star formation properties}
\begin{figure}
\center
\includegraphics[angle=0,scale=.8]{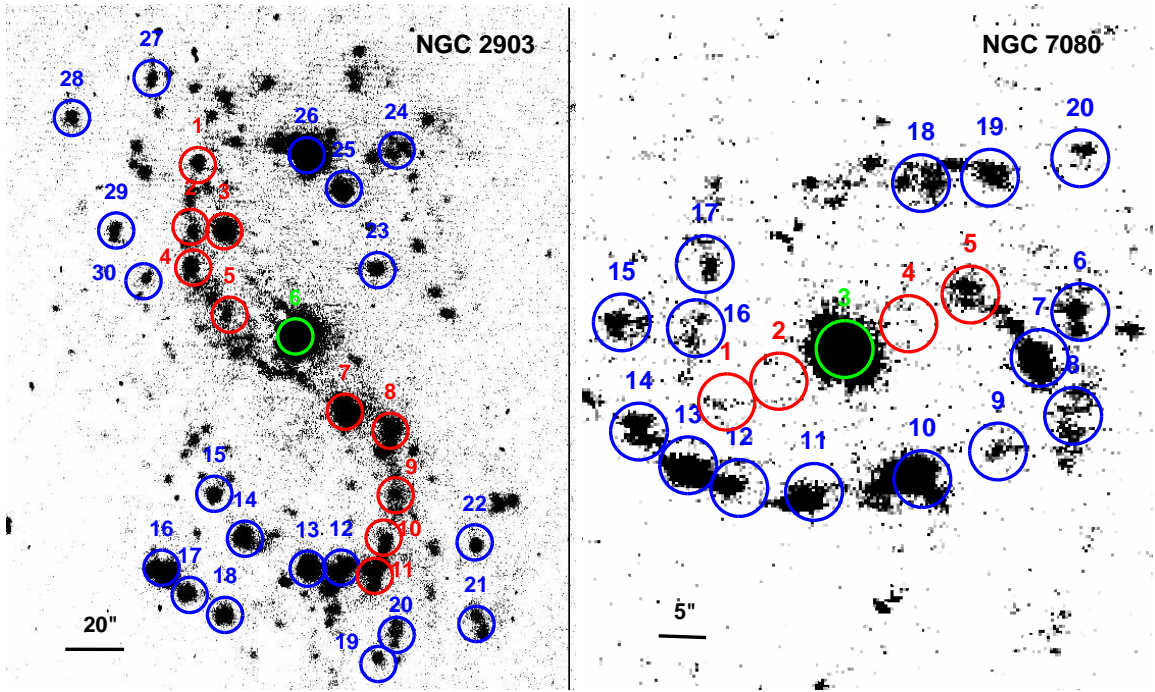}
\caption{Continuum-subtracted H$\alpha$ images of NGC 2903 ({\it left}) and NGC 7080 ({\it right}) with the star formation clumps identified. The measuring apertures are 6$''$ (336 pc) for NGC 2903, and 3$''$ (885 pc) for NGC 7080. The clumps in the bulge regions are marked with green circles, the clumps in the bar regions are marked with red circles, and disk clumps are marked with blue circles.
\label{fig5}}
\end{figure}

To study the properties of the star formation activities in different galactic structural regions, we identified 30 and 20 clumps by visual inspection of the H$\alpha$ images in NGC 2903 and NGC 7080, respectively (Figure ~\ref{fig5}). To strengthen our detection criterion, we required the clumps to be also the emission peaks in the MIPS 24 $\mu$m images and the apertures to be non-overlapping, although some fainter clumps were excluded, because their regions were strongly overlapped, when they were very close to bright clumps. However, even no clear emissions can be found in the bar region of NGC 7080, four clumps (clump 1, 2, 4 and 5) were still marked for comparison with other regions. The IRAF task {\it phot} in the DAOPHOT package was used to performed aperture photometry in the H$\alpha$ and IR images of both galaxies. For the clumps in NGC 2903, we used circular apertures of 6$''$ radii, corresponding to $\sim$ 336 pc at the galaxy distance. For the clumps in NGC 7080, the apertures of 3$''$ radii were used, corresponding to $\sim$ 885 pc.
The measured fluxes of the clumps are listed in Table \ref{Table3}. This table lists the clump ID in Column
(1), the right ascension and declination columns (2) and (3), and also marked the structures the clumps located in Column (4). Columns (5) shows the H$\alpha$ narrowband fluxes, and columns (6)--(10) show the IR fluxes.

In our imaging bands, H$\alpha$, 8 $\mu$m(dust), and 24 $\mu$m emissions are all important tracers of star formation activities, which have been has been confirmed in many studies \citep[e.g.,][]{Kennicutt98, Calzetti05, Wu05, Zhu08, Fisher09}. Considering the spatial resolution of our data, we used the H$\alpha$ and 8 $\mu$m(dust) emissions as the star formation tracers, and derived the star formation rates (SFRs) of the clumps following the equation given by \citet{Kennicutt09} (their Eq.[11] and Table [4]):
\begin{equation}
SFR_{H\alpha +8\mu m}(M_{\odot}yr^{-1})=7.9\times 10^{-42}[L(H\alpha)_{obs}+0.010L(8\mu m (dust))](erg~s^{-1}).
\end{equation}
In addition, the 3.6 $\mu$m band is often treated as a good tracer for galactic stellar mass owing to its depth and its reddening-free sensitivity mainly to older stars \citep[e.g.,][]{Li07, Sheth10, Zhu10}. Thus, we calculated the stellar mass of the clumps based on the formula derived by \citet{Zhu10}:
\begin{equation}
Log {~M(M_{\odot})}=(-0.79\pm0.03)+(1.19\pm0.01)\times Log { ~\nu L_{\nu}{[3.6\mu m]}(L_{\odot})}.
\end{equation}

\begin{figure}
\center
\includegraphics[angle=0,scale=.8]{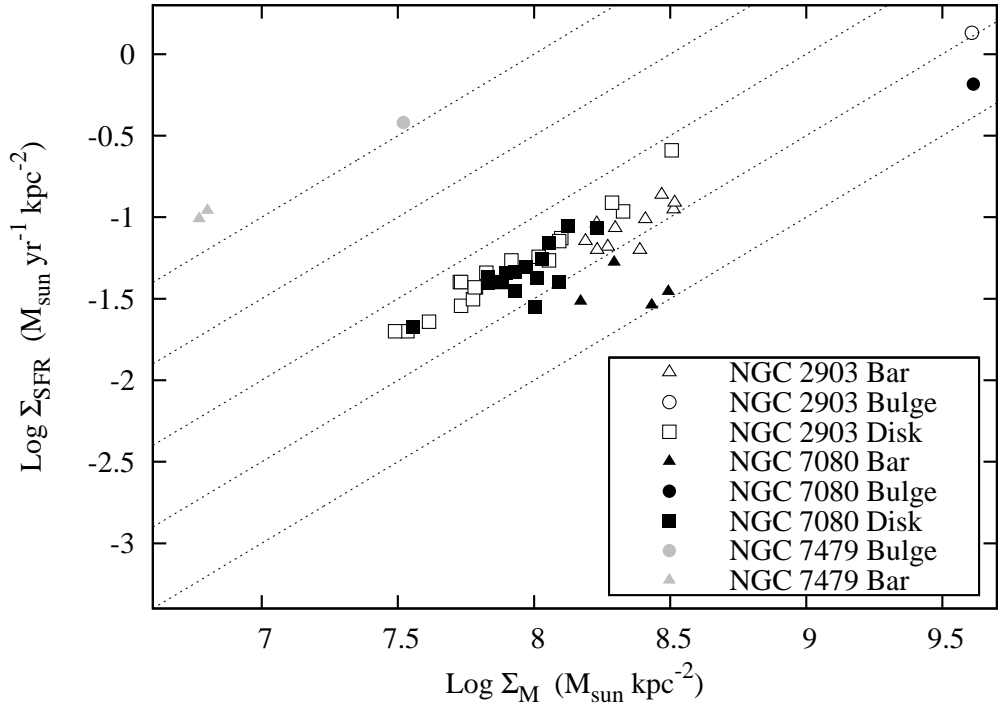}
\caption{Relation between the SFR surface density and stellar mass surface density for the clumps in NGC 2903 and in NGC 7080. The open symbols represent the clumps in NGC 2903, and the solid ones represent the clumps in NGC 7080. The bulge and stellar bar of NGC 7479 (gray symbols) from Paper I are also plotted for comparison. The dotted lines show constant timescales that the present-day SFR needs to form the entire stellar mass. The timescales are (increasing downward) of 10$^8$, 10$^{8.5}$, 10$^9$, 10$^{9.5}$, and 10$^{10}$ yr.
\label{fig6}}
\end{figure}

The SFR surface density of the clumps in NGC 2903 and in NGC 7080 is plotted in Figure \ref{fig6} against the stellar mass surface density within the clumps. The symbols of circles, triangles and squares represent the clumps in the bulges, bars and disks, respectively. NGC 2903 has solid symbols and NGC 7080 has filled symbols. Typically, the stellar mass surface densities are similar when the clumps are located in the same structures in both galaxies, while the clumps in NGC 2903 have higher SFR surface densities. The SFR surface densities in the two bulge clumps are 1.35 and 0.65 M$_{\odot}$ yr$^{-1}$ kpc$^{-2}$ for NGC 2903 and 7080, respectively, and are higher by one order of magnitude than those in the bars and disks. Except the bulges, the bar clumps in NGC 2903 have higher SFR surface densities than most of the disk clumps. The bar clumps in NGC 7080, on the other hand, have lower SFR surface densities, although their stellar mass surface densities are higher than those in the disks. In this figure, we also plotted the values of NGC 7479 from Paper I for comparison. The SFR surface density of the bulge in NGC 7479 is lower than both bulges of the other two galaxies, although the SFR surface densities of the bar in NGC 7479 are similar with those in the bar of NGC 2903. Similarly, the bulge and bar of NGC 7479 also have lower stellar mass surface densities. In addition, there are less significant differences between the bulge and bar of NGC 7479 in their values. In NGC 7479, the surface densities of the bulge are $\sim$ 0.5 dex higher than those of the bar, while they are $\sim$ 1 dex in NGC 2903 and NGC 7080. However, it should be noted that the actual surface densities of the bulge and bar in NGC 7479 should be higher due to the background contamination in the photometry.

The same diagram can be used to constrain the timescale on which the stellar mass is built in the present SFR. Figure \ref{fig6} shows lines of constant stellar mass growth times of 10$^8$, 10$^{8.5}$, 10$^9$, 10$^{9.5}$, and 10$^{10}$ yr. We found that nearly all disk clumps in both galaxies can grow up in the present SFR in 1--3 Gyr. The clumps in the bulge and the bar of NGC 2903 have a timescale of $\sim$ 3 Gyr, a little longer than their associated disk. While all the bulge and bar clumps in NGC 7080 have a timescale longer than 3 Gyr, especially for the two bar clumps nearest the bulge, whose timescale is nearly 10 Gyr. This indicates that the present star formation activities in the bar of NGC 7080 contribute little to the building of the stellar bar. Although the timescale for the bulge and bar of NGC 7479 is shorter than 0.1 Gyr, it is likely the result of its large photometric aperture and another driving factor beside the stellar bar affecting the star formation activities (Paper I).

%%Sect.5 Discussion: Bar-driven Secular Evolution
\section{Bar-driven Secular Evolution}
The internal secular processes play an important role in galaxy evolution. Especially as the Universe expands, the secular evolution becomes more and more important, and will finally become dominant, overtaking mergers \citep{KK04}. By comparing the relative efficiency of gas inflows based on SFR and gas-phase metallicity, \citet{Ellison11} estimated that bars contribute at least 3 times more to the centrally triggered star formation than the external effect of galaxy-galaxy interactions. Numerous studies of the dynamics in barred galaxies have suggested that under the influence of the bar, the gas inside the corotation undergoes negative gravity torques, and loose angular momentum in a rapid central gas inflow towards the central kpc or less \citep{Athanassoula92, Sellwood93}.

Barred galaxies such as NGC 2903 and 7080 provide an ideal laboratory to study the bar-driven secular evolution, in which star formation may take place in the galactic central regions as the result of gas inflow. To measure the ability of driving gas inflow, the parameters of bars have been investigated by many researchers, both physical and dynamical \citep[e.g.,][]{Laurikainen04, Gadotti07, Marinova07}. As shown in previous statistics \citep{Barazza08, Gadotti11}, the local bars have diameters of 2 to 24 kpc, with a mean of 2--6 kpc, and have the normalized size to the associated disk (r$_{e,bar}$/R$_{24}$) of 0.2--0.4. The effective radii of the bars in NGC 2903 and NGC 7080 are $>$5 kpc and $>$3 kpc, respectively (Sec. 3.2). This may indicate that the two galaxies have long stellar bars relatively, although their relative size r$_{e,bar}$/R$_{25}$ are similar with most local bars\citep{Barazza08}. In addition, the two bars also have high ellipticities ({\it e}$_{bar}$ = 0.70 and 0.88, for NGC 7080 and NGC 2903 respectively) compared with other local bars of which the majority have ellipticities 0.50 $\leq$ {\it e}$_{bar}$ $\leq$ 0.75 \citep{Marinova07}.

The effect of bars on the galaxy evolution have been confirmed by a myriad of observations, including the central molecular gas concentrations \citep[e.g.,][]{Sakamoto99, Komugi08} and star formation activities \citep[e.g.,][]{Roussel01, Shi06, Mazzuca08}. As a result of the secular evolution, bar redistribution of mass and the vertical resonance can build spheroids in the galactic central regions \citep{Combes90}, which are called pseudobulges. Many studies have indicated that the properties of pseudobulges, different from the merger-built bulges, are linked to those of their associated outer disks \citep[e.g.,][]{Laurikainen07, Mazzuca08}. Based on these properties, we confirmed that the bulges of NGC 2903 and NGC 7080 are both pseudobulges (Sec. 3.2). In the high-resolution optical image from HST WFPC2 (Figure~\ref{fig1}), the dust structures are clearly revealed in the central region of NGC 2903, which confirms the identification of its pseudobulge. Based both on bulge morphology and S\'ersic index, \citet{Fisher09} also classified NGC 2903 as having a pseudobulge. In general, pseudobulges have the SFR densities of 0.1--1 M$_{odot}$ yr$^{-1}$ kpc$^{-2}$, higher than the galactic disks \citep{KK04}, consistent with what we found in Figure~\ref{fig6}. In addition, in the pseudobulges of our investigated galaxies, the present-day SFR needs a few Gyrs to form their entire stellar mass. This is also consistent with the result timescale of 2--8 Gyr in \citet{Fisher09}.

To better understand the evolutionary sequence of bar-driven secular evolution, several scenarios have been proposed \citep[e.g.,][]{Martin97, Verley07}. \citet{Jogee05} analyzed the potential evolutionary connections between the barred starbursts and non-starbursts, and projected a figure of the dynamical evolution driven by stellar bars. In this evolutionary figure, the evolutionary sequence of a strongly barred galaxy was divided into three stages: type I non-starburst, type II non-starburst, and circumnuclear starburst. Type I non-starburst is the early stages of bar-driven inflow, with gas still inflowing along the bar and not forming stars efficiently. In the type II stage, most of the gas has piled up in the circumnuclear region, and then there is a circumnuclear starburst when most of the gas exceeds a critical density. Based on the scenario of \citet{Jogee05}, \citet{Sheth05} supplemented a category, type III non-starburst. This stage is in the poststarburst phase, where the gas has been consumed by a circumnuclear starburst, and no molecular gas is within the bar region. We compared our results with this evolutionary sequence. There are intense star formation activities both in the central regions of NGC 2903 and NGC 7080. However, unlike NGC 7479 which is the type I non-starburst with clear star formation regions located in the stellar bar (Paper I), only a few star formation clumps in the bar of NGC 2903 can be identified, with the properties similar to those in the disks, rather than the bulge (Figure~\ref{fig6}).  Therefore, we suggested that, NGC 2903 is likely in the late stage of the type II non-starburst, where sufficient gas has accumulated to drive a starburst in the central region, although some gas is still in the bar. Under this scenario, more intense star formation activity may take place in the circumnuclear region of NGC 2903 later, along with the star formation activities disappearing in the galactic bar. The properties of the molecular gas also support our interpretation. Based on HCN(1--0) emission in NGC 2903, \citet{Leon08} found that the dynamical time for the inflow of gas, from the bar to the center, is considerably shorter than the consumption time scale in the bar, indicating that most of the gas in the bar will be concentrated in the center of this galaxy, although some other process\footnote{Processes such as minor merger, galaxy tidal encounters, and long-timescale external gas accretion likely also lead to this result.} may also contribute to this result \citep{Popping10}.

For NGC 7080, the SFR densities of the clumps in the bar are even lower than the disk clumps, the most intense star formation activity takes place in its bulge region, along with clumpy H{\sc ii} knots located in the spiral arms. It is probably that nearly all of the gas in the bar of NGC 7080 has been concentrated in the bulge region. Therefore this galaxy is likely in the stage of the circumnuclear starburst, which will evolve to the type III non-starburst phase after the gas in the central region has been consumed.

%%Sect.6 Summary
\section{Summary}
In this study, we presented {\it Spitzer} mid-IR and ground-based optical images of two barred galaxies NGC 2903 and NGC 7080. Based on the images, we studied the global and structural properties of the two galaxies, analyzed the star formation activities in their bulges, bars and disks, and investigated the internal secular evolution driven by their stellar bars.

We performed 2D three-component bulge-disk-bar decomposition with the data analysis algorithm GALFIT based on the IRAC 3.6 $\mu$m images, and obtained the parameters of each galactic structure. The bulges of the two galaxies both have low S\'ersic indices (n $<$ 1), along with low bulge-to-total luminosity ratios. As compared with previous statistics, it is suggested that the two bulges are both pseudobulges, the products of the galaxy secular evolution. The stellar bars of the two galaxies are of large physical sizes and high ellipticities compared with the bars in the local Universe.

We identified 30 and 20 star formation clumps in NGC 2903 and NGC 7080, respectively. These clumps were located in the galactic bulges, bars and disks. The two bulge clumps have the highest SFR surface densities, which are higher by one order of magnitude than those in the bars and disks. In the present SFR, the clumps in the bulge and the bar of NGC 2903 need $\sim$ 3Gyr to form its entire stellar mass, a little longer than the clumps in the disks, while the same structural clumps in NGC 7080 need a much longer time.

Finally, we compared our results with the possible scenario of bar-driven secular evolution proposed by \citet{Jogee05}, and suggested that NGC 2903 is likely in the late stage of the type II non-starburst. NGC 7080 is likely in the stage of the circumnuclear starburst until the gas in the central region is consumed.

\begin{acknowledgements}
We sincerely thank the anonymous referee for his/her very careful review of this paper, constructive advice, and very useful suggestions. This project is supported by NSFC grants 11173030, 10833006, 10773014, 10978014 and 11078017 and the 973 Program grant 2007CB815406 and 2012CB821803.

This work was partially supported by the Open Project Program of the Key Laboratory of Optical Astronomy, National Astronomical Observatories, Chinese Academy of Sciences. This work is based on observations made with the {\it Spitzer Space Telescope}. The {\it Spitzer Space Telescope} is operated by the Jet Propulsion Laboratory, California Institute of Technology, under NASA contract 1407. We made extensive use of the NASA/IPAC Extragalactic Database (NED) which is operated by the Jet Propulsion Laboratory, California Institute of Technology, under contract with the National Aeronautics and Space Administration.
\end{acknowledgements}

\clearpage

%%Table 1 %%

\begin{table}
\bc
\begin{minipage}[]{75mm}
\caption[]{The basic parameters of the galaxies}\label{Table1}
\end{minipage}
\small
\begin{tabular}{lll}
  \hline\noalign{\smallskip}
  \hline\noalign{\smallskip}
 Parameter &  NGC 2903  &  NGC 7080 \\
  \hline\noalign{\smallskip}
R.A. (J2000.0)$^a$ & 09h32m10.1s & 21h30m01.9s \\
Dec. (J2000.0)$^a$ & 21d30m03s & 26d43m04s \\
Classification & SB(r)d & SB(r)b \\
V$_{hel}$ (km s$^{−1}$)$^a$ & 556 & 4839 \\
Distance (Mpc)$^b$ & 8.9 & 62.7 \\
Linear scale (pc arcsec$^{-1}$) & 56 & 295 \\
D$_{25}$$^c$ & 12.6$'$ & 1.8$'$\\
Major axis P.A.$^c$ & $17^\circ$ & $20^\circ$ \\
{\it e}$^d$ & 0.52 & 0.05 \\
m$_B$$^c$ & 9.59 & 13.73 \\
  \noalign{\smallskip}\hline
\end{tabular}
\ec
\tablecomments{0.5\textwidth}{\\
$^a$ From NED\\
$^b$ The luminosity distance\\
$^c$ From RC3 \citep{de91}\\
$^d$ The ellipticity of the isophotes at 25 B mag/arcsec$^2$}

\end{table}

%%Table 2%%%
\begin{table}
\bc
\begin{minipage}[]{70mm}
\caption[]{Parameters From 2D Decomposition}\label{Table2}
\end{minipage}
\small
\begin{tabular}{ccccccccc}
  \hline\noalign{\smallskip}
  \hline\noalign{\smallskip}
Galaxy Name & Bulge n & Bulge r$_e$ (kpc) & Bar r$_e$ (kpc) & r$_{e,bar}$/R$_{25}$ & {\it e}$_{bar}$ & Disk r$_{s}$ (kpc) & B/T (\%)& Bar/T (\%)\\
(1) & (2) & (3) & (4) & (5) & (6) & (7) & (8) & (9) \\
  \hline\noalign{\smallskip}
NGC 2903 (Original)      & 0.50 & 0.32 & 3.51 & 0.16 & 0.88 & 2.90 & 7.66 & 6.35 \\
NGC 2903 (PAH-corrected) & 0.58 & 0.33 & 3.36 & 0.15 & 0.85 & 2.94 & 4.53 & 6.91 \\
  \\
NGC 7080 (Original)      & 0.68 & 0.61 & 5.21 & 0.32 & 0.70 & 4.61 & 18.72 & 11.03 \\
NGC 7080 (PAH-corrected) & 0.63 & 0.53 & 3.77 & 0.23 & 0.55 & 5.07 & 16.18 & 11.31 \\
  \noalign{\smallskip}\hline
\end{tabular}
\ec
\tablecomments{1.2\textwidth}{Columns are: (1) Galaxy Name. (2) Bulge S\'ersic index. (3) Bulge effective radius in kpc. (4) Bar effective radius in kpc. (5) Size of the bar relative to disc radius R$_{25}$ (6) Bar ellipticity. (7) Disk scale-length in kpc. (8) Bulge-to-total light ratio. (9) Bar-to-total light ratio. The original results were derived based on the original 3.6 $\mu$m images. The PAH-corrected ones were derived based on the contaminant-corrected 3.6 $\mu$m images, which were obtained following the method in \citet{Kendall11}.}

\end{table}

%%Table 3%%%
\begin{table}
\bc
\begin{minipage}[]{80mm}
\caption[]{Positions and Photometry of the Clumps}\label{Table3}
\end{minipage}
\small
\begin{tabular}{ccccrrrrrr}
  \hline\noalign{\smallskip}
  \hline\noalign{\smallskip}
ID & R.A.$^{a}$ & Dec.$^{a}$ & Position & H$\alpha$$^{b}$ & 3.6 $\mu$m$^{c}$ & 4.5 $\mu$m$^{c}$ & 5.8 $\mu$m$^{c}$ & 8 $\mu$m(dust)$^{c}$ & 24 $\mu$m$^{c}$\\
  \hline\noalign{\smallskip}
\multicolumn{10}{c}{NGC 2903}\\
  \hline\noalign{\smallskip}
1 & 09:32:12 & +21:31:04 & bar & 26.13 & 13.03 & 8.29 & 23.37 & 57.79 & 48.15\\
2 & 09:32:12 & +21:30:43 & bar & 21.29 & 8.13 & 5.18 & 14.36 & 33.94 & 31.61\\
3 & 09:32:12 & +21:30:41 & bar & 60.70 & 11.94 & 8.39 & 27.50 & 69.29 & 72.04\\
4 & 09:32:12 & +21:30:29 & bar & 31.95 & 7.54 & 4.84 & 13.49 & 31.23 & 28.95\\
5 & 09:32:12 & +21:30:13 & bar & 22.76 & 10.21 & 6.59 & 15.03 & 31.96 & 34.85\\
6 & 09:32:10 & +21:30:05 & bulge & 367.05 & 108.16 & 75.92 & 272.68 & 704.14 & 1123.63\\
7 & 09:32:09 & +21:29:40 & bar & 66.34 & 13.07 & 8.82 & 24.30 & 60.21 & 52.32\\
8 & 09:32:08 & +21:29:34 & bar & 41.26 & 6.95 & 4.63 & 13.82 & 34.96 & 29.36\\
9 & 09:32:07 & +21:29:12 & bar & 22.48 & 8.57 & 5.69 & 17.59 & 45.54 & 35.28\\
10 & 09:32:08 & +21:28:57 & bar & 26.23 & 10.63 & 6.90 & 19.80 & 50.54 & 45.69\\
11 & 09:32:08 & +21:28:44 & bar & 47.01 & 7.52 & 4.83 & 17.57 & 45.71 & 46.11\\
12 & 09:32:09 & +21:28:47 & disk & 53.52 & 9.06 & 6.03 & 21.11 & 54.46 & 56.44\\
13 & 09:32:10 & +21:28:47 & disk & 81.01 & 8.39 & 5.79 & 22.00 & 57.98 & 84.23\\
14 & 09:32:11 & +21:28:57 & disk & 50.19 & 5.85 & 4.04 & 13.92 & 34.86 & 46.81\\
15 & 09:32:12 & +21:29:10 & disk & 26.23 & 4.98 & 3.34 & 11.48 & 28.88 & 35.81\\
16 & 09:32:13 & +21:28:47 & disk & 64.37 & 3.43 & 2.42 & 7.86 & 18.41 & 22.71\\
17 & 09:32:12 & +21:28:38 & disk & 40.89 & 2.84 & 1.98 & 7.02 & 17.34 & 28.94\\
18 & 09:32:12 & +21:28:31 & disk & 22.57 & 3.19 & 2.09 & 7.34 & 17.64 & 12.70\\
19 & 09:32:08 & +21:28:14 & disk & 11.60 & 2.87 & 1.82 & 5.32 & 13.83 & 11.72\\
20 & 09:32:07 & +21:28:24 & disk & 17.21 & 3.12 & 2.00 & 5.82 & 15.77 & 10.47\\
21 & 09:32:05 & +21:28:28 & disk & 22.60 & 1.96 & 1.31 & 3.51 & 8.18 & 7.03\\
22 & 09:32:06 & +21:28:56 & disk & 19.66 & 2.28 & 1.51 & 4.43 & 10.47 & 10.11\\
23 & 09:32:08 & +21:30:28 & disk & 22.06 & 5.36 & 3.58 & 10.96 & 26.79 & 19.69\\
24 & 09:32:07 & +21:31:09 & disk & 25.33 & 4.10 & 2.82 & 11.20 & 27.76 & 21.19\\
25 & 09:32:09 & +21:30:56 & disk & 43.23 & 5.76 & 3.93 & 14.08 & 35.01 & 33.78\\
26 & 09:32:10 & +21:31:07 & disk & 138.15 & 12.78 & 9.85 & 46.74 & 125.54 & 160.81\\
27 & 09:32:13 & +21:31:34 & disk & 15.39 & 2.87 & 1.97 & 8.14 & 21.00 & 16.40\\
28 & 09:32:15 & +21:31:20 & disk & 16.04 & 1.80 & 1.22 & 3.88 & 8.95 & 7.39\\
29 & 09:32:14 & +21:30:42 & disk & 14.23 & 3.16 & 2.14 & 7.65 & 18.95 & 27.95\\
30 & 09:32:14 & +21:30:24 & disk & 12.83 & 3.80 & 2.49 & 8.46 & 20.50 & 21.21\\
  \hline\noalign{\smallskip}
\multicolumn{10}{c}{NGC 7080}\\
  \hline\noalign{\smallskip}
1 & 21:30:03 & +26:42:59 & bar & 2.78 & 0.70 & 0.44 & 0.99 & 2.05 & 2.41\\
2$^d$ & 21:30:02 & +26:43:01 & bar & 2.48 & 1.15 & 0.74 & 1.05 & 1.97 & 1.38$^e$\\
3 & 21:30:02 & +26:43:05 & bulge & 84.06 & 11.36 & 7.61 & 17.14 & 41.35 & 53.06\\
4$^d$ & 21:30:01 & +26:43:07 & bar &3.21 & 1.30 & 0.80 & 1.27 & 2.37 & 1.72$^e$\\
5 & 21:30:01 & +26:43:11 & bar & 7.28 & 0.88 & 0.56 & 1.41 & 3.27 & 2.96\\
6 & 21:30:00 & +26:43:09 & disk & 7.53 & 0.36 & 0.24 & 1.00 & 2.46 & 1.98\\
7 & 21:30:00 & +26:43:04 & disk & 14.04 & 0.78 & 0.53 & 1.97 & 5.06 & 4.06\\
8 & 21:30:00 & +26:42:58 & disk & 7.44 & 0.44 & 0.29 & 1.06 & 2.72 & 1.99\\
9 & 21:30:01 & +26:42:54 & disk & 4.54 & 0.50 & 0.33 & 0.80 & 1.67 & 1.53\\
10 & 21:30:01 & +26:42:51 & disk & 24.73 & 0.64 & 0.45 & 1.59 & 3.96 & 3.23\\
11 & 21:30:02 & +26:42:50 & disk & 9.45 & 0.51 & 0.36 & 1.00 & 2.19 & 1.98\\
12 & 21:30:03 & +26:42:50 & disk & 7.87 & 0.53 & 0.36 & 1.34 & 3.39 & 2.69\\
13 & 21:30:03 & +26:42:52 & disk & 14.31 & 0.55 & 0.39 & 1.51 & 3.78 & 2.99\\
14 & 21:30:04 & +26:42:56 & disk & 8.11 & 0.41 & 0.28 & 1.04 & 2.61 & 2.00\\
15 & 21:30:04 & +26:43:08 & disk & 6.91 & 0.40 & 0.27 & 0.95 & 2.30 & 1.95\\
16 & 21:30:03 & +26:43:07 & disk & 5.27 & 0.60 & 0.39 & 1.07 & 2.51 & 2.17\\
17 & 21:30:03 & +26:43:14 & disk & 5.21 & 0.44 & 0.28 & 0.90 & 2.16 & 1.81\\
18 & 21:30:01 & +26:43:22 & disk & 8.00 & 0.47 & 0.30 & 1.20 & 2.91 & 1.86\\
19 & 21:30:01 & +26:43:23 & disk & 6.60 & 0.36 & 0.23 & 0.97 & 2.31 & 1.52\\
20 & 21:30:00 & +26:43:25 & disk & 4.20 & 0.21 & 0.15 & 0.56 & 1.16 & 0.78\\
  \noalign{\smallskip}\hline
\end{tabular}
\ec
\tablecomments{1.2\textwidth}{\\
$^a$ Units of right ascension are hours, minutes, and seconds, and units of declination are degrees, arcminutes, and arcseconds.\\
$^b$ In units of 10$^{-15}\ erg\ s^{-1}\ cm^{-2}$.\\
$^c$ In units of mJy; 1 mJy = 10$^{-26}\ erg\ s^{-1}\ cm^{-2}\ Hz^{-1}$.\\
$^d$ The two clumps were selected mainly because of their position in the bar regions, although there were no clear H$\alpha$ emission in them. \\
$^e$ The clumps were contaminated by the diffraction of the galactic bulge in the 24$\mu$m image. Therefore, the values were calculated using the 8.0$\mu$m flux based on Equation (8) in \citet{Calzetti05}.}

\end{table}

%%Figture%%%

\end{document}